\documentstyle[twoside,fleqn,npb,epsfig]{article}
\newcommand{\AmS}{{\protect\the\textfont2
  A\kern-.1667em\lower.5ex\hbox{M}\kern-.125emS}}

\title{Exclusive $\rho^{0}$ Production in Polarized DIS at SMC}

\author{Ang\`ele Tripet for the Spin Muon Collaboration \\
        Fakult\"at f\"ur Physik, Universit\"at Bielefeld, 
        33501 Bielefeld, Germany\\
        E-mail: Angele.Tripet@cern.ch}

\begin{document}

\begin{abstract}
Preliminary spin cross section asymmetries for exclusive $\rho^{0}$ 
lepto-production, $ \vec{\mu} + \vec{N} \rightarrow \mu^{\prime} + N + \rho^{0} $ 
$ (\rho^{0} \rightarrow \pi^{+} \pi^{-})$, are reported. 
These asymmetries have been determined 	for the first time 
by the Spin Muon Collaboration (SMC) 
at low $Q^{2}$ (photoproduction) and at large $Q^{2}$ (DIS) 
for different $p_{T}^{2}$ intervals in the kinematic range 
0.01~$<Q^{2}<$~60~GeV$^{2}$ 
and 140~$<$~W$^{2}$~$<$~310~GeV$^{2}$  ($<W> \simeq $~15~GeV)
for the full SMC data set. About 100 K $\rho^{0}$'s have been 
selected for 0.62~$<$~$m ( \pi^{+} \pi^{-} ) $~$<$~1.07~GeV$/c^{2}$ 
and $ |I|< 0.05$. Within the statistical precision, no significant asymmetries 
have been observed at low $Q^{2}$ in the preliminary results.
\end{abstract}

\maketitle

\section{INTRODUCTION}

For the first time, results on exclusive $\rho^{0}$ 
spin cross section asymmetries have been presented in this workshop. 
The experiment was carried out at CERN by the Spin Muon Collaboration 
(SMC). SMC is a fixed target experiment using a 190 GeV polarized muon 
beam impinging on polarized proton (butanol or ammonia) and deuteron 
(deu\-te\-ra\-ted butanol) targets in the kinematic range 
0.0008~$<$~x~$<$~0.7, 0.01~GeV$^{2}$~$< Q^{2} <$~60~GeV$^{2}$, and 
140~$<$~W$^{2}$~$<$~310~GeV$^{2}$  ($<W> \simeq $~15~GeV).
A detailed description of the SMC experiment is 
given in ref \cite{SMC97}.
Only aspects particular to the exclusive 
$\rho^{0}$ production will be discussed.

Exclusive $\rho^{0}$ production is a process in which 
only a $\rho^{0}$ is produced and the nucleon recoils elastically: 
$ \vec{\mu} + \vec{N} \rightarrow \mu^{\prime} + N + \rho^{0}$~
$ (\rho^{0} \rightarrow \pi^{+} \pi^{-})$. 
In SMC the kinematics is entirely determined by the 
scattered muon, which is precisely reconstructed even at low $Q^{2}$, 
and the $\rho^{0}$ via its decay into $\pi^{+}$~and~$\pi^{-}$. 
Although the nucleon is undetected, the kinematics of the scattered muon and of 
the $\rho^{0}$ are sufficient to select the elastic process \cite{NMC94}.

Such a measurement will add more insight into the 
$\rho^{0}$ production mechanism, as well as into its spin dependence. 
It will also allow to study the spin properties of the hadronic photon at very low $Q^{2}$. 
Given the high energy of the SMC muon beam,  
the spin cross section asymmetries are 
measured in the $<W>$ region of $\sim$~15 GeV, 
where at low $Q^{2}$ unpolarized data are interpreted in terms of 
a {\it soft} pomeron exchange~\cite{POM}. 

\section{THE $\rho^{0}$ EVENT SAMPLE}

First, {\it good} DIS events are selected based on the scattered 
muon kinematics: {\bf $Q^{2} >$ 0.01 GeV$^{2}$}, {\bf $y >$ 0.4}  and {\bf $y <$ 0.9}, 
{\bf scattered muon momentum $>$ 19 GeV}. 

\begin{figure}[t]
\begin{center}
\vspace{-1.0cm}
\mbox{\epsfig{figure=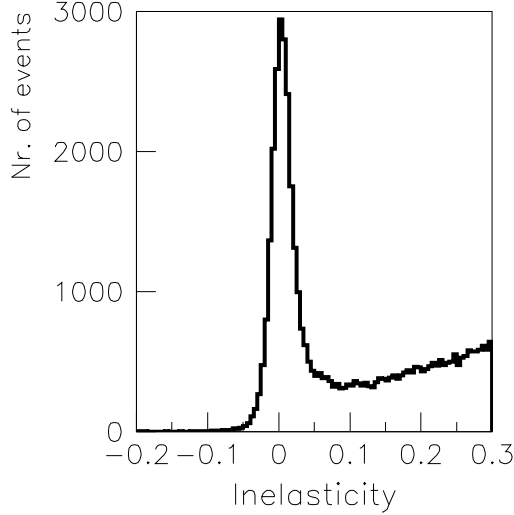,width=0.43\textwidth}}
\vspace{-0.9cm}
\caption{Inelasticity distribution (1995 data) after selections.}
\label{figure1}
\vspace{-1.0cm}
\end{center}
\end{figure}

Second, the vector mesons are identified via their 
decay into $\pi^{+}$ and $\pi^{-}$. In order to obtain an elastic 
$\rho^{0}$ sample, events have been selected requiring: \\
a) {\bf only two hadron tracks of opposite charge} associated 
with the vertex defined by the incident and scattered muon tracks. 
To each track the pion mass has been assigned, since there was no 
hadron identification. 
Electron pairs from brems\-strahlung photons have been removed 
u\-sing the SMC calorimeter \cite{PRETZ}, and 
events with the invariant mass  $m ( K^{+} K^{-} ) $ in the range of the $\phi$ meson 
have been excluded. \\
b) to isolate an elastic process, events in the inelasticity, I, interval 
{\bf $-$0.05~$<$~I~$<$~0.05}, where 
I~$= (M_{x}^{2} - M_{p}^{2}) / W^{2}$ ($M_{x}$ is 
the missing mass squared of the undetected recoiling system), have been selected. 
Fi\-gure \ref{figure1} shows the inelasticity distribution. 
The peak around zero contains the elastic events above a very small background. 
Other experiments cut on  
$\Delta E  = (M_{x}^{2} - M_{p}^{2}) / 2 M_{p}$ $( \simeq$ 
I $\cdot \nu$ for large $\nu )$. In SMC the width of the $\Delta E$ distribution 
is about $m_{\pi}$, and the I interval corresponds to 
$-$~0.5~$<$~$\Delta E $~$<$~0.5~GeV. The I cut has been 
chosen because it does not depend on $W^{2}$; \\
c) finally a cut on the invariant mass has been applied:
{\bf 0.62~$<$~$m ( \pi^{+} \pi^{-} ) $~$<$~1.07}~GeV$/c^{2}$.

\begin{figure}[htb]
\begin{center}
\vspace{-1.0cm}
\mbox{\epsfig{figure=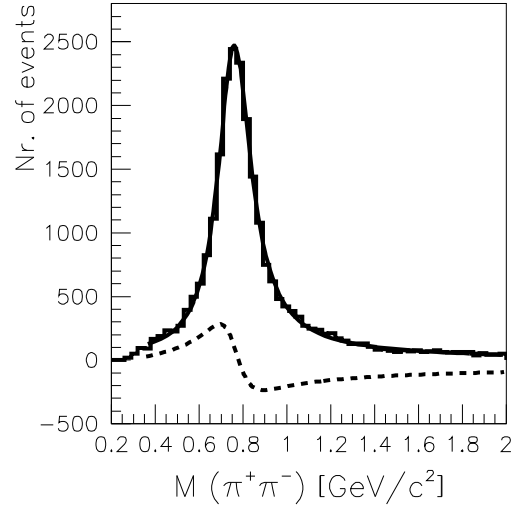,width=0.43\textwidth}}
\vspace{-0.9cm}
\caption{Mass spectrum (1995 data) after selections. The full line 
represents a fit according to the S\"oding model, the dotted line the inter\-fe\-rence 
term between the resonant and non resonant amplitudes.}
\label{figure2}
\vspace{-1.0cm}
\end{center}
\end{figure}

Figure~\ref{figure2} shows the invariant mass spectrum 
of the selected data sample with a clear $\rho^{0}$ peak. This  
$\rho^{0}$ invariant mass distribution is well described by 
different models. In Figure~\ref{figure2} a fit to the distribution 
is performed according to the S\"oding model \cite{SOED}.

The final $\rho^{0}$ sample consists of $\sim$~52~K~$\rho^{0}$'s produced on polarized protons and 
$\sim$~43~K~$\rho^{0}$'s on polarized deuterons.
For $p_{T}^{2}>~0.09$~GeV$^{2}/c^{2}$, $\sim$~17~K~$\rho^{0}$'s on 
protons and $\sim$~14~K~$\rho^{0}$'s on deuterons 
are left, where $p_{T}$ is the $\rho^{0}$ transverse momentum 
with respect to the virtual photon.

\section{THE SPIN CROSS SECTION ASYMMETRIES}

The $\rho^{0}$ lepton-nucleon spin cross section asymmetry 
is given by: 
\begin{eqnarray}
A^{l N \rightarrow \rho^{0} l^{\prime} N}_{LL}
 =  \frac{\Delta \sigma_{\parallel}}{2 \overline{\sigma}} 
 =   \frac{\sigma^{lN} (\Uparrow \downarrow) - \sigma^{lN} (\Uparrow \uparrow)}
{\sigma^{lN} (\Uparrow \downarrow) + \sigma^{lN} (\Uparrow \uparrow)} =  \nonumber \\
=  \frac{1}{f} \cdot
\frac{1}{P_{b}} \cdot \frac{1}{P_{t}} \cdot \frac{1}{2} [
\frac{N_{\Uparrow \downarrow} - N_{\Uparrow \uparrow}}
     {N_{\Uparrow \downarrow} + N_{\Uparrow \uparrow}} -
\frac{N_{\Uparrow \downarrow}^{\prime} - N_{\Uparrow \uparrow}^{\prime}}
     {N_{\Uparrow \downarrow}^{\prime} + N_{\Uparrow \uparrow}^{\prime}}
  ]   
\nonumber
\end{eqnarray}
 
\noindent 
where $N$ ($N^{\prime}$) is the number of reconstructed events 
before (after) the target polarization reversal.
The indices $\Uparrow \downarrow$ and $\Uparrow \uparrow$ refer to 
the re\-la\-tive orientation of the photon and proton (or deuteron) spins.  
$P_{t}$ and $P_{b}$ are the target and beam 
polarizations, respectively, and 
\begin{eqnarray}
\nonumber f = \frac{n_{p(d)} \sigma_{p(d)}}
         {n_{p(d)} \sigma_{p(d)} + \Sigma_{A} n_{A} \sigma_{A}} \nonumber
\end{eqnarray}
\noindent
is the dilution factor giving the fraction of po\-la\-ri\-zed nucleons to the 
total nucleons in the target. 
In order to estimate $f=f$($Q^{2}$, p$_{T}^{2}$), all available cross 
section measurements for elastic $\rho^{0}$ pro\-du\-ction on nuclei 
in our kinematic range were used.

\begin{figure}[htb]
\begin{center}
\vspace{-1.7cm}
\mbox{\epsfig{figure=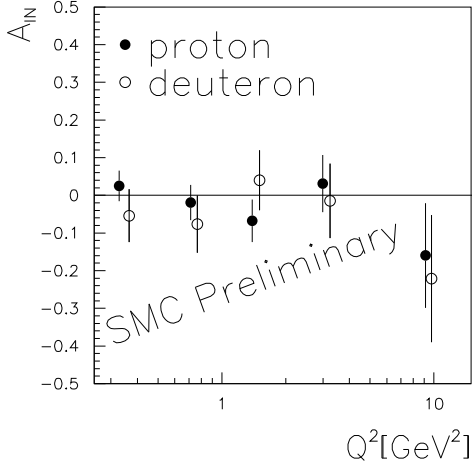,width=0.47\textwidth}}
\vspace{-1.7cm}
\caption{$A^{l N \rightarrow \rho^{0} l^{\prime} N}_{LL}$ for proton 
and deuteron as a function of $Q^{2}$ (the errors are statistical only).}
\label{figure3}
\vspace{-1.0cm}
\end{center}
\end{figure}
\begin{figure}[htb]
\begin{center}
\vspace{-1.1cm}
\mbox{\epsfig{figure=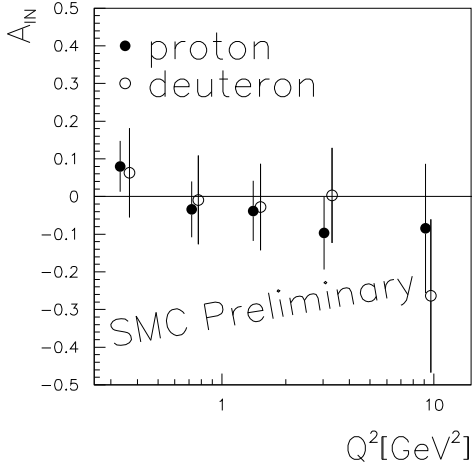,width=0.47\textwidth}}
\vspace{-1.7cm}
\caption{$A^{l N \rightarrow \rho^{0} l^{\prime} N}_{LL}$  
for $p_{T}^{2} > $~0.09~GeV$^{2}/c^{2}$ 
(the errors are statistical only).}
\label{figure4}
\vspace{-1.1cm}
\end{center}
\end{figure}

Figure~\ref{figure3} shows the $\rho^{0}$ production 
spin asymmetries for the proton and the deuteron separately. The data have 
been divided in 5 $Q^{2}$ bins in order to study the $Q^{2}$ dependence. 
In Figure~\ref{figure4} the same data are shown for 
$p_{T}^{2} > 0.09$ GeV$^{2}/c^{2}$. In both cases the data 
show no significant asymmetry neither for the proton 
nor for the deuteron.  Note that the spin  
asymmetry originates only from scattering on the polarized 
proton or deuteron, while the $\rho^{0}$'s produced from nuclei 
represent an unpolarized background, which is properly taken into 
account by the dilution factor.

\begin{figure}[t]
\begin{center}
\vspace{-1.73cm}
\mbox{\epsfig{figure=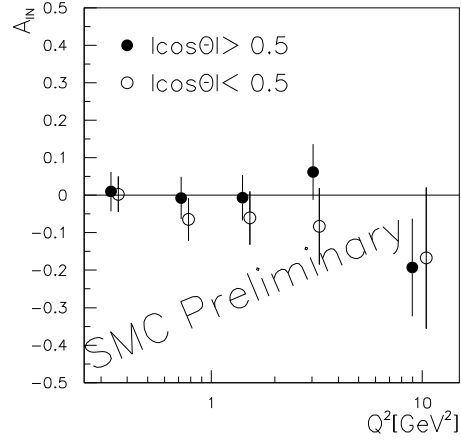,width=0.485\textwidth}}
\vspace{-2.1cm}
\caption{$A^{l N \rightarrow \rho^{0} l^{\prime} N}_{LL}$ for  
$| \cos \, \vartheta | > $~0.5 and $| \cos \, \vartheta | < $~0.5 for proton 
and deuteron combined (the errors are statistical only).}
\label{figure5}
\vspace{-1.1cm}
\end{center}
\end{figure}

The $\rho^{0}$ lepton-nucleon production spin asymmetries have also been studied as a function of 
the $\vartheta$ polar angle of the decay $\pi^{+}$ in the 
$\rho^{0}$~c.m.s. For $| \cos \, \vartheta | < $ 0.5 ($| \cos \, \vartheta | > $ 0.5) 
the $\rho^{0}$'s  are mainly transversely (longitudinally) polarized. 
Figure~\ref{figure5} shows 
$A^{l N \rightarrow \rho^{0} l^{\prime} N}_{LL}$ 
as a function of 
$Q^{2}$ for  $| \cos \, \vartheta | > $~0.5 and  $| \cos \, \vartheta | < $~0.5 
(proton and deuteron combined).

As a check of systematic effects, the false asymmetries, 
which by construction should 
give zero, have been computed. They have been found to be zero. 
The contribution of radiative events has been estimated to be 
less than 2~\% for the SMC kinematics.

The main sources of systematic errors are \\
1) the non-elastic background below the inelasticity peak: it is 
less than 5~\% at low $Q^{2}$, and slightly higher (less than 10~\%) at high $Q^{2}$; \\
2) the uncertainty on the dilution factor co\-ming from 
the poor knowledge of the A-de\-pen\-dence of the cross section 
$\sigma_{A} (Q^{2}, p_{T}^{2})$ giving 
$ \frac{\Delta f}{f} < $ 0.15.

\vspace*{-0.3cm}
\section{SUMMARY}

The SMC collaboration at CERN has presented for the first time 
preliminary results on exclusive spin cross section asymmetries 
of the $\rho^{0}$ mesons 
for $Q ^{2} >$~0.01~GeV$^{2}$ and $<W> \, \simeq $~15~GeV 
over a large $Q ^{2}$ range.
The preliminary data, within the statistical accuracy of the measurement,  
show no significant spin asymmetries 
at low $Q^{2}$ ($Q^{2} <$ 5~GeV$^{2}$)
both for the whole $p_{T}^{2}$ range 
and $p_{T}^{2} > 0.09$~GeV$^{2}/c^{2}$, neither for the proton nor for the deuteron. 
In the last $Q^{2}$ bin ($Q^{2} \sim$ 10~GeV$^{2}$) an indication of a 
non-zero spin asymmetry 
is observed. Combining proton and deuteron data, a value of -0.18~$\pm$~0.11 
for $A^{l N \rightarrow \rho^{0} l^{\prime} N}_{LL}$ is obtained.

\nopagebreak[4]

\end{document}